\makeatletter
\declare@file@substitution{revtex4-1.cls}{revtex4-2.cls}
\makeatother
\documentclass[onecolumn]{aastex631}
\usepackage{textcase}
\shorttitle{4U 1626$-$67}
\shortauthors{Zhou et al.}
\graphicspath{{./}{figures/}}
\begin{document}

\title{QPOs in a highly magnetized ultra-compact X-ray binary 4U 1626$-$67}

\author[0009-0009-7509-3794]{Zi-Yi Zhou}
\affiliation{School of Physics and Astronomy, Sun Yat-sen University, Zhuhai, 519082, People’s Republic of China}

\author[0000-0001-9599-7285]{Long Ji$^*$}
\affiliation{School of Physics and Astronomy, Sun Yat-sen University, Zhuhai, 519082, People’s Republic of China}
\affiliation{CSST Science Center for the Guangdong-Hong Kong-Macau Greater Bay Area, DaXue Road 2, 519082, Zhuhai, People's Republic of China}
\email{jilong@mail.sysu.edu.cn}

\author[0000-0003-3188-9079]{Ling-Da Kong}
\affiliation{Institute f{\"u}r Astronomie und Astrophysik, Kepler Center for Astro and Particle Physics, Eberhard Karls, Universit{\"a}t, Sand 1, D-72076 T{\"u}bingen, Germany}

\author[0000-0002-9679-0793]{Sergey S. Tsygankov}
\affiliation{Department of Physics and Astronomy, FI-20014 University of Turku, Finland}

\author[0000-0001-5160-3344]{Qing-Cang Shui}
\affiliation{Key Laboratory of Particle Astrophysics, Institute of High Energy Physics, Chinese Academy of Sciences, 100049, Beijing, China}
\affiliation{University of Chinese Academy of Sciences, Chinese Academy of Sciences, 100049, Beijing, China}

\author[0000-0002-2705-4338]{Lian Tao}
\affiliation{Key Laboratory of Particle Astrophysics, Institute of High Energy Physics, Chinese Academy of Sciences, 100049, Beijing, China}


\correspondingauthor{Long Ji}


\begin{abstract}

We report the detection of mHz quasi-periodic oscillations (QPOs) in four \textit{NuSTAR} observations of 4U 1626$–$67 during its recent spin-down episode.
By using a novel method based on the Hilbert-Huang Transform (HHT), we present the first QPO-phase-resolved timing and spectral analysis of accreting X-ray pulsars in low mass X-ray binaries.
Broadband QPO waveforms have been reconstructed and exhibit approximately sinusoidal shapes, with fractional amplitudes that vary with energy.
In addition, we find that spin pulse profiles exhibit stable shapes between different QPO phases with different instantaneous fluxes, while the fractional root-mean-square (rms) is distinct for different observations. 
In this source, both QPO-phase-resolved and averaged spectra can be modeled with a negative and positive powerlaws exponential (NPEX) model, and their spectral evolutions show a similar trend,
suggesting that the QPO modulation is caused by accretion rate variability instead of a geometric obscuration.
These results provide new constraints on accretion physics in strongly magnetized neutron stars and the underlying mechanisms of QPOs.

\end{abstract}

\keywords{accretion, accretion discs-stars: neutron-pulsars: individual: 4U 1626$-$67 -X-rays: binaries.}


\section{Introduction} \label{sec:1}

4U 1626$-$67 is an ultra-compact low mass X-ray binary that harbors an accreting X-ray pulsar with a spin period of $\sim$ 7.7 seconds and a low mass companion of $\lesssim$ 0.02\,$M_{\odot}$
\citep{Middleditch..1981ApJ...244.1001M, 
Chakrabarty..1997ApJ...474..414C, Li..1980ApJ...240..628L}.
Since its discovery by {\it Uhuru} in 1972 \citep{Giacconi..1972ApJ...178..281G}, it has undergone several torque reversals, switching between spin-up and spin-down episodes, accompanied by significant changes in luminosity \citep[e.g.,][]{Camero..2010ApJ...708.1500C, Charkrabarty1997}.
Notably, during the spin-down phase, 4U 1626$-$67 exhibits prominent QPOs around 47\,mHz in the power density spectra, which however are less pronounced during the spin-up phase \citep{Chakrabarty..1998ApJ...492..342C, Jain..2010MNRAS.403..920J}. 
%
QPOs are non-coherent signals that are widely observed in X-ray binaries across different types of compact objects, including white dwarfs, neutron stars and black holes, although their underlying mechanisms might be distinct \citep[e.g.,][]{Warner2004,Wijnands1999,Manikantan2024,Ingram2019}.
In accreting X-ray pulsars, the physical origin of QPOs is still unclear.
\citet{van..1987ApJ...316..411V} suggested that QPOs might be attributed to the obscuration by blobs in the inner part of the accretion disk.
In this case, the QPO frequency corresponds to the Keplerian frequency.
Alternatively, \citet{Alpar..1985Natur.316..239A} proposed a beat frequency model, which suggests that the spatial configuration of the neutron star's magnetic field will change as matter flows inward from the accretion disk, resulting in fluctuations in the accretion rate and producing the observed QPO phenomenon. 
It predicts that the QPO frequency equals to the difference between the spin frequency and the Keplerian frequency at the Alfv\'en radius.

The broadband spectra of 4U 1626$-$67 have been studied by different authors \citep[e.g.,][]{Sharma..2023MNRAS.526L..35S, Camero..2012A&A...546A..40C}, and its spectral shape during the spin-down period is somewhat softer than that during the spin-up period \citep{Yi1999}. 
A cyclotron resonance scattering feature (CRSF) around 37\,keV has been detected in both spin-up and spin-down phases, suggesting its magnetic field of $\sim 4\times10^{12}$\,G \citep{Orlandini..1998ApJ...500L.163O,D'A..2017MNRAS.470.2457D}.
Previous studies predominantly focus on the time-averaged spectra,
and investigations regarding QPO-modulated emission is relatively rare.
This is due to the difficulty of determining the QPO phase
given that the QPO frequency is variable, unlike coherent periodic signals.
Recently, a method, based on the variational mode decomposition (VMD) and the Hilbert-Huang transform, has been developed and successfully applied to QPOs in black hole systems \citep{Shui2023,Shui2024a,Shui2024b,Shui2024c}.
This technique allows for the adaptive decomposition of signals into independent components, enabling the separation of QPO modulations from other variabilities.
In this paper, we apply this method to QPOs in a highly magnetized X-ray pulsar 4U 1626$-$67 and perform QPO-phase-resolved studies.


\begin{table}
        \centering
        \caption{Summary of {\it NuSTAR} observations. 
        Two rate columns give the background-subtracted count rates combining FPMA and FPMB modules in the energy ranges of {3-10}\,keV and {10-79}\,keV for each observation. The last two columns $K$ and \textit{a} are parameters used for performing the variational mode decomposition.}
    \begin{tabular}{cccccccc}
        \toprule
         ObsID & {Date} & Exposure  & Rate ({3-10}\,keV) & Rate ({10-79}\,keV) & $K$ & \textit{a} \\
        &(yyyy-mm-dd) & (s) & cts/s & cts/s & & \\
        \hline
         90901318002 & 2023-05-02 & 27422 & $4.18\pm0.01$ & $2.33\pm0.01$ & 5 & 1000 \\
         90901318006 & 2023-06-04 & 18468 & $3.40\pm0.02$ & $1.90\pm0.01$ & 6 & 900 \\
         90901318008 & 2023-06-22 & 22332 & $3.31\pm0.01$ & $1.68\pm0.01$ & 5 & 1500 \\
         90901318010 & 2023-07-05 & 18449 & $2.98\pm0.01$ & $1.55\pm0.01$ & 5 & 1000 \\
        \hline
    \end{tabular}
    \label{tab:observations}
\end{table}

\section{OBSERVATIONS AND DATA REDUCTION} \label{sec:2}

{\it Nuclear Spectroscopic Telescope Array} ({\it NuSTAR}) is a space-based X-ray observatory launched by NASA in 2012, capable of observing X-rays in the energy range of 3 to 79 keV. 
Equipped with two focal plane modules (FPMA and FPMB) which are operated simultaneously, {\it NuSTAR} can provide high resolution imaging and spectroscopy \citep{Harrison..2013ApJ...770..103H}. 
In this paper, we studied four {\it NuSTAR} observations of 4U 1626$-$67 carried out during its recent spin-down episode when QPOs appear (Table~\ref{tab:observations}). 

Data reduction was carried out using the NuSTAR Data Analysis Software in {\sc HEASOFT} (version 6.32.1) with {\tt CALDB} (version 20240520). 
We calibrated and screened the event files using the {\tt NUPIPELINE} task, and employed the {\tt NUPRODUCTS} task to extract lightcurves and spectra. The source and background spectra were extracted from circular regions of equal radius (80 arcsec). The source region was centered on the source position, while the background region was placed away from the source.
All lightcurves in the following were background-subtracted after combining FPMA and FPMB modules. During spectral analysis, we grouped the spectra using the {\tt FTGROUPPHA} task. We followed the optimal binning scheme proposed by \citet{ftgrouppha..2016A&A...587A.151K} with more than 25 counts in each bin using {\tt grouptype=optmin} and {\tt groupscale=25}. All uncertainties quoted in this paper correspond to a confidence level of 90$\%$.

\section{RESULTS} \label{sec:3}
\subsection{Decomposition of QPO signals}\label{subsec:3.1}
\begin{figure*}[t!]
    \centering
    \includegraphics[width=0.8\textwidth]{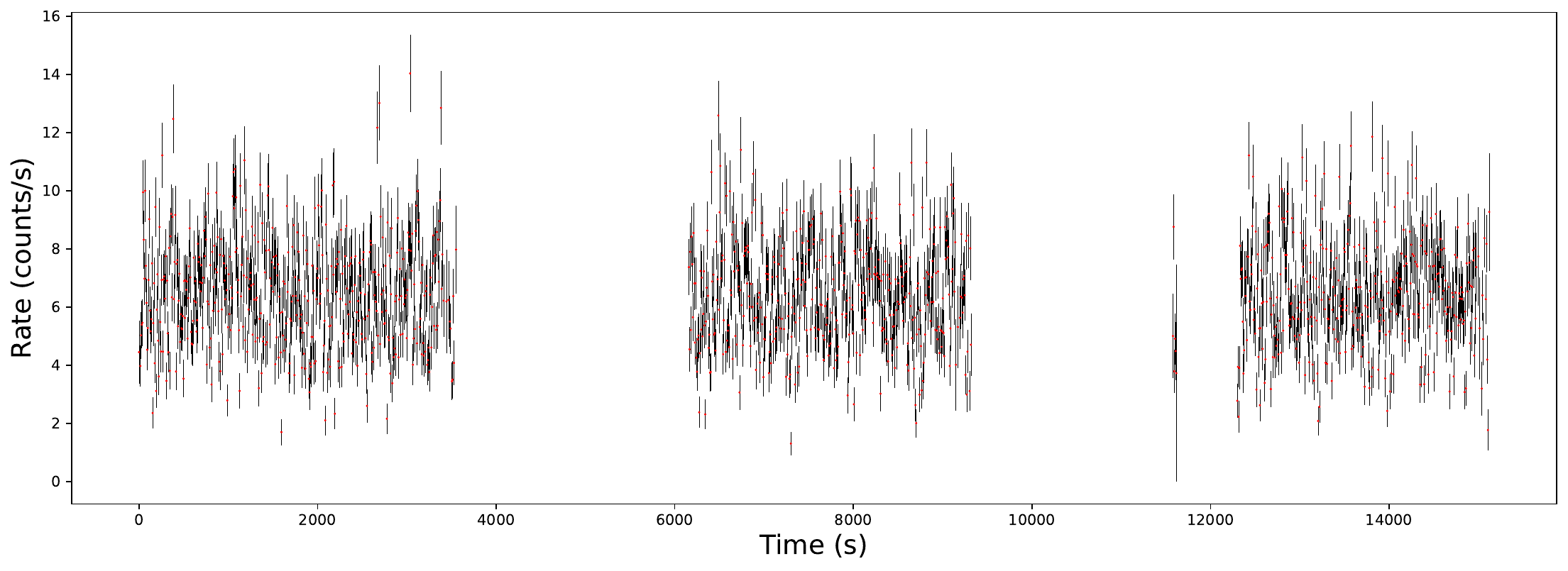}
    \caption{A representative segment of the lightcurve for the obsID 90901318002 with a binsize 7.7\,s.}
    \label{fig:lc}
\end{figure*}

\begin{figure*}[t!]
    \centering
    \includegraphics[width=0.4\textwidth]{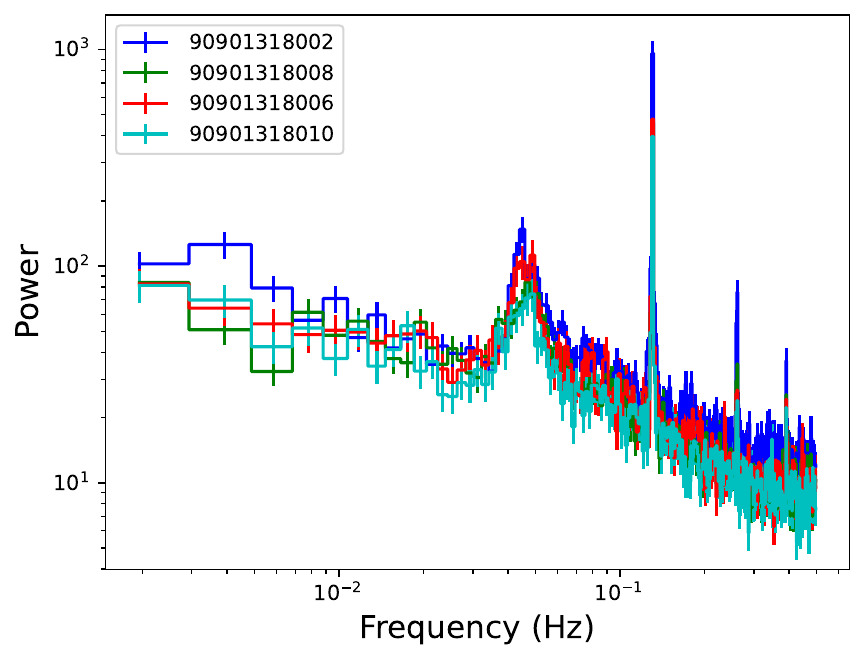}
    \hspace{0.1cm}
    \includegraphics[width=0.4\textwidth]{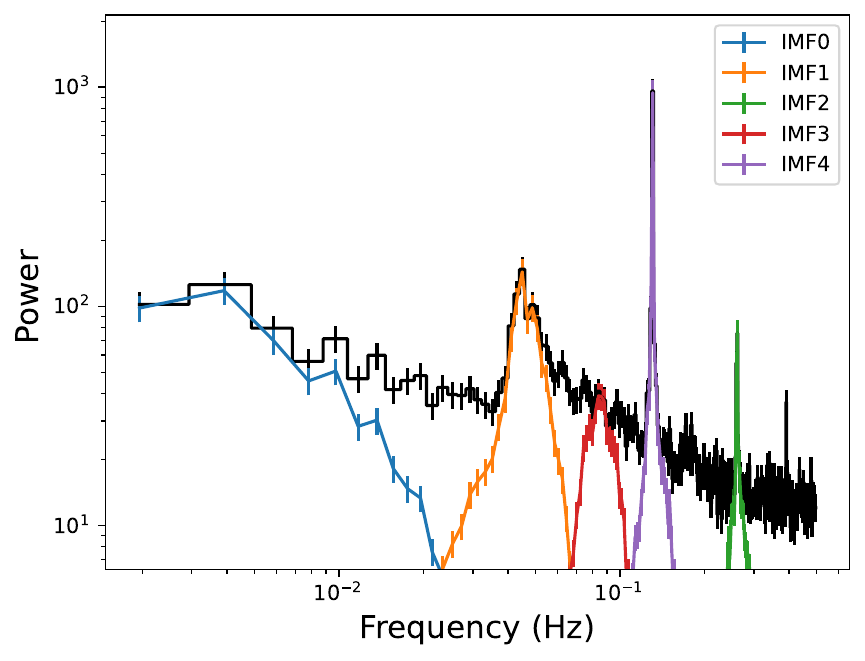}
    \caption{Left: power density spectra (PDS) of four {\it NuSTAR} observations of 4U 1626$-$67. Right: a representative PDS (black line) constructed for the observational ID 90901318002, together with PDSs of individual IMFs after the decomposition. 
    The IMF1 (yellow line) corresponds to the QPO signal we considered.
    Here we assumed $K$=5 and \textit{a}=1000 (see text).
    }
    \label{fig:PDS}
\end{figure*}

We first inspected overall lightcurves for each observation with a bin size of 7.7\,s (i.e., the period of the pulsar), and found that they remain generally stable although with intrinsic variations (Figure~\ref{fig:lc}).
Therefore, we treated each observation as a whole to perform the timing analysis.
We then extracted lightcurves in the energy range of 4-79\,keV with a bin size of 1\,s and calculated their power spectra using a Python package {\tt Stingray}\footnote{\url{https://github.com/StingraySoftware/stingray/tree/v1.1.2}} \citep{Huppenkothen..2019ApJ...881...39H} in the frequency range of 0.002-0.5\,Hz.
We show the results in Figure~\ref{fig:PDS}.
Generally, the four power spectra exhibit a similar shape, which presents a strong QPO around 47\,mHz, consistent with previous papers \citep{Sharma..2023MNRAS.526L..35S, Tobrej..2024MNRAS.528.3550T, 4U162667QPO..2025MNRAS.538.1046S}. 

%
%
%
%

We used the Variational Mode Decomposition (VMD) technique to 
decompose the original lightcurves into a set of intrinsic mode functions (IMFs) and each IMF corresponds to variations of a narrow frequency band (see Appendix \ref{sec:appendix A}). 
This method has been successfully used to separate QPO signals in black holes X-ray binaries \citep[e.g.,][]{Shui2023}, and we followed the same approach here.
We performed this calculation using an open source Python package Vmdpy\footnote{\url{https://github.com/vrcarva/vmdpy}} (for details, see Appendix \ref{sec:appendix A}; \citet{VMDPY..CARVALHO2020102073}).
During the decomposition, there are two parameters $K$ and \textit{a}, where $K$ means the number of independent components and both $K$ and \textit{a} influence the frequency width in the power spectrum.
Following \citet{Shui..2023ApJ...957...84S}, we determined the combination of $K$ and \textit{a} by comparing the power spectra of the resulting IMFs with that of the original lightcurve.
In practice, we initially set a trial value of $K$ with variable \textit{a}, and incremented it until the QPO component could be separated.
We found that $K$=5 is generally the minimum value to identify  coherent pulsations, the QPO component and its harmonic unambiguously. 
After determining $K$, we adjusted the value of \textit{a} to make sure that one IMF's PDS can match the QPO of the original lightcurve.
In Figure~\ref{fig:PDS}, we show an example of the decomposition, where the IMF1 represents the QPO component.
In Table~\ref{tab:observations}, we list the combination of $K$ and \textit{a} used in the paper.
We note that although the selection of $K$ and \textit{a} is artificial, as shown in Appendix~\ref{sec:appendix B}, it has little influence on the reconstructed QPO signals.

\subsection{QPO waveforms}
\begin{figure}[h!]
\centering
\includegraphics[width=0.45\textwidth]{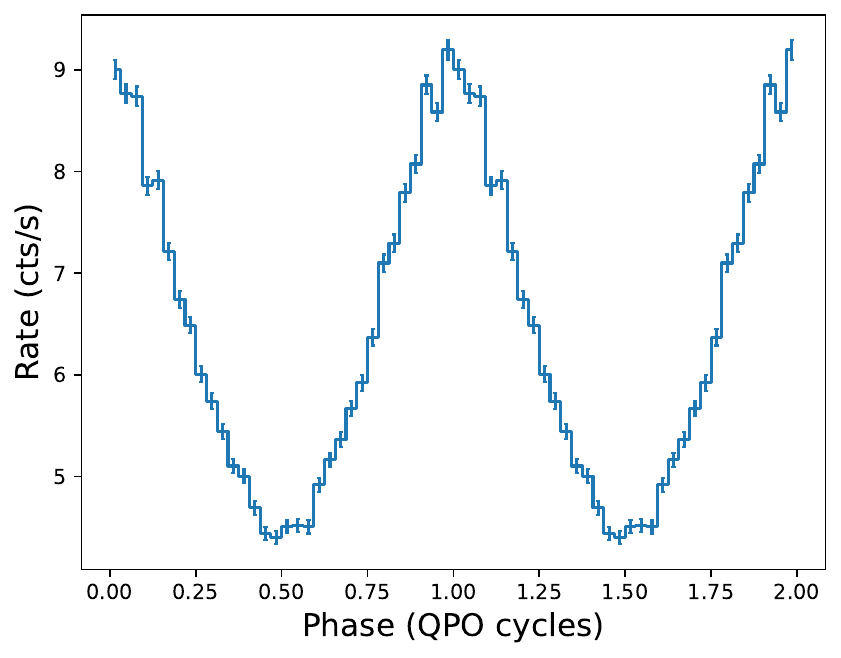}
\caption{The QPO waveform reconstructed by folding the 4-79\,keV lightcurve according to the instantaneous phase obtaining from the HHT using the data from ObsID 90901318002.
Here we divide the QPO cycles into 32 phase bins and present two QPO cycles for clarity.}

\label{fig:002profile}
\end{figure}

\begin{figure}
	\centering
	\includegraphics[width=0.45\textwidth]{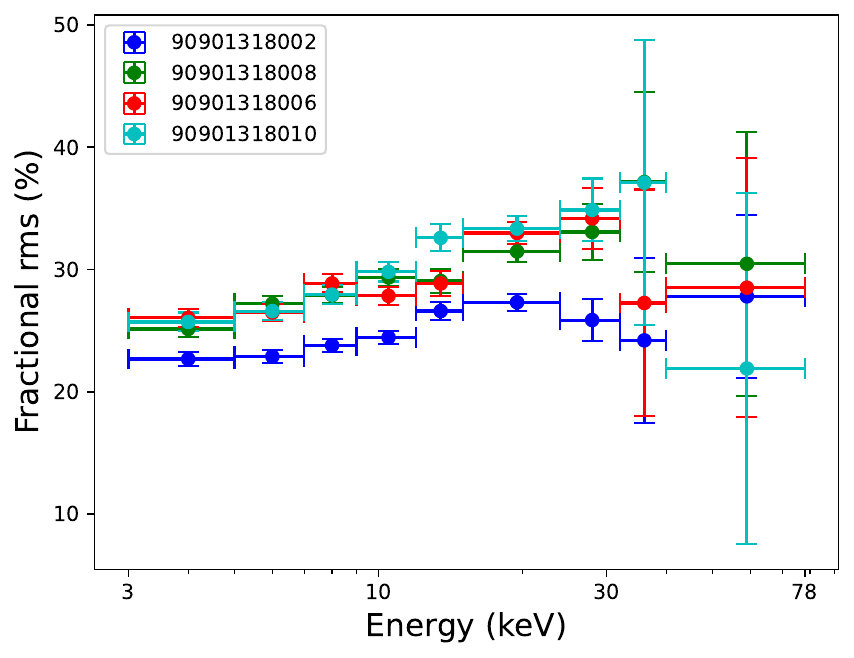}
	\caption{The evolution of the fractional rms of QPO waveforms with energy in four observations of 4U 1626$-$67.
    }
	\label{fig:energy-resolved}
\end{figure}

Once the IMF that corresponds to the QPO component had been decomposed, we could determine its instantaneous QPO phase for a given time through the Hilbert transform \citep[][]{Shui..2024ApJ...965L...7S, Huang..1998RSPSA.454..903H}.
Then we folded the original lightcurve according to the resulting QPO phase to obtain QPO waveforms.
In Figure~\ref{fig:002profile}, we present an example using the data from the observational ID 90901318002.
Using QPO phases obtained above, energy-dependent QPO waveforms can be constructed as well.
In this paper, we selected energy bands of 3-5\,keV, 5-7\,keV, 7-9\,keV, 9-12\,keV, 12-15\,keV, 15-24\,keV, 24-32\,keV, 32-40\,keV, and 40-78\,keV.
Generally, QPO waveforms in different energy bands have a similar shape.
To quantify the amplitude of modulation, we calculated their fractional rms\footnote{For a comparison with the conventional method, see Appendix~\ref{sec:appendix C}.} following \citet{Dhillon2009}
\begin{equation}
    rms = \frac{\sqrt{\sum_{i = 0}^{N} \left[ (p_i - \bar{p})^2 - \sigma_{p_i}^2 \right] / N}}{\bar{p}},
\label{PF}
\end{equation}
where $\bar p$ is the average count rate, $p_i$ and $\sigma_{p_i}$ are the count rate and its error for the phase bin $i$ of the QPO waveform, respectively. 
As shown in Figure~\ref{fig:energy-resolved}, the fractional rms increases with energy in all four observations, i.e., QPOs are more significant for hard X-rays.
In addition, we found that this increasing trend seems to be flatten above 15\,keV and there are large error bars above 30\,keV due to poor statistics. 

\subsection{Spin pulse profiles at different QPO phases}
\begin{figure*}[h!]
    \centering
    \includegraphics[width=0.48\linewidth]{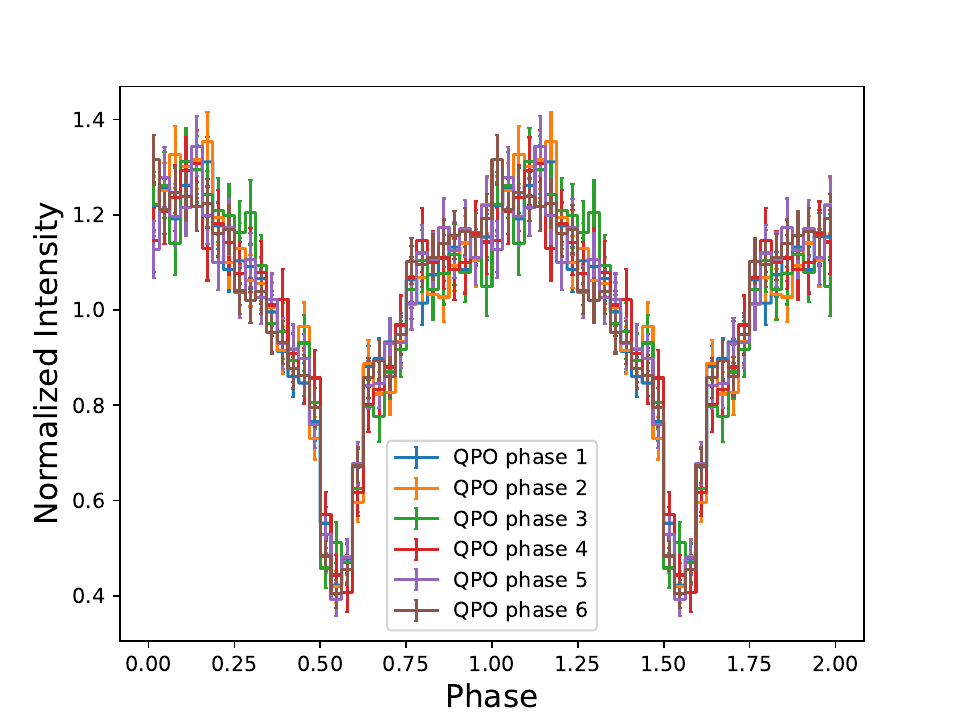}
    \includegraphics[width=0.48\linewidth]{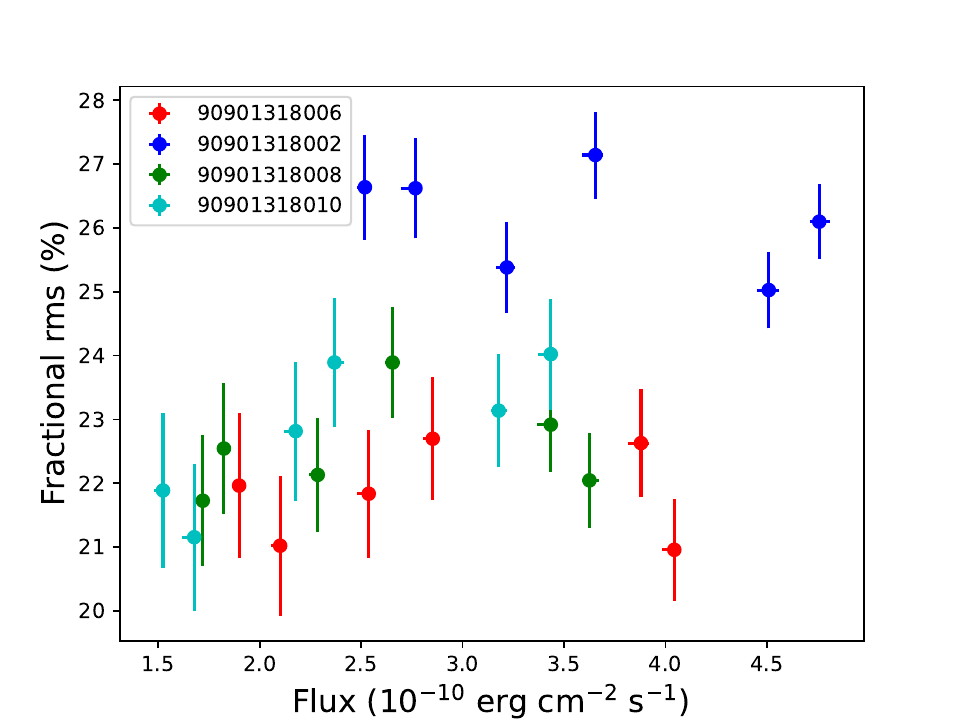}
    \caption{
%
Left: normalized pulse profiles of 4U 1626$-$67 at six equally spaced QPO phases (labeled 1–6) from ObsID 9090131808. Right: fractional rms of pulse profiles measured at different QPO phases across observations, as a function of the observed flux.
    }
    \label{fig:flux_pulsed_fraction}
\end{figure*}

As suggested by \citet{Kommers1998}, QPOs can modulate the instantaneous amplitude of coherent pulsations, as indicated by the detection of sidebands in power spectra.
Therefore, in this study, we explored the variation of spin pulse profiles at different QPO phases.
In practice, we first estimated the spin period for each observation using the {\tt EFSEARCH}\footnote{\url{https://heasarc.gsfc.nasa.gov/xanadu/xronos/help/efsearch.html}} task.
Then we folded the lightcurves extracted from six QPO phases as defined in Section~\ref{subsec:3.2}. 
We found that although the absolute amplitudes of pulse profiles are modulated by QPOs, their shapes are very similar across different QPO phases, regardless of the instantaneous fluxes (Figure~\ref{fig:flux_pulsed_fraction}).
Moreover, when comparing the pulse profiles across different observations, the pulsed fractions (see the right panel of Figure~\ref{fig:flux_pulsed_fraction}), calculated using Eq.~\ref{PF}, were notably different, even for some QPO phases with similar flux levels.


\subsection{Spectral analysis\label{subsec:3.2}}
The X-ray continuum spectrum of 4U 1626$-$67 can be described by either a negative and positive powerlaws exponential (NPEX) model or a combination of a blackbody model and a powerlaw model \citep[e.g.,][]{Sharma..2023MNRAS.526L..35S, Camero..2012A&A...546A..40C}.
In this work, we adopted the NPEX model, i.e.,
\begin{equation}
F(E)=  N( f E^{-\alpha} + E^{+\beta}) \exp( -\frac{E}{E_{\rm cut}}),
\label{eq:npex}
\end{equation}
where the positive power index $\beta$ was fixed to 2, representing the Wien portion of the thermal distribution \citep{Mihara..1995PhDT.......215M}.
The cyclotron resonant scattering feature (CRSF) has been reported around 37\,keV, and we modeled it using a $cyclabs$ component following \citet[][]{D'A..2017MNRAS.470.2457D}.
Above 60\,keV, there might be a CRSF harmonic, although the detection is not significant due to limited statistics \citep{Coburn..2002ApJ...580..394C}.
As suggested by \citet{Sharma..2023MNRAS.526L..35S}, we included another absorption component \textit{gabs} to describe this harmonic, and fixed its centroid energy and width at 65.9\,keV and 9\,keV, respectively.
To account for the Fe K$\alpha$ fluorescence line, we employed a \textit{Gaussian} emission line model and fixed its energy and width to 6.4\,keV and 0.1\,keV, respectively.
We also used the \textit{tbabs} model \citep{Wilms2000} for the photoionization absorption of the interstellar medium, and the equivalent hydrogen ($N_{\rm H}$) was fixed at $0.096\times10^{22}\,\rm cm^{-2}$\citep{HI4PI..2016A&A...594A.116H}.
In addition, a \textit{const} model was used to adjust the inner-calibration between instruments and we set the FPMA value to 1.
We found that this model could well describe the average spectra for all four observations.
The best-fit parameters are listed in Table~\ref{tab:fitting para}.


\begin{figure*}[t!]
    \centering
    \includegraphics[width=0.230\textwidth]{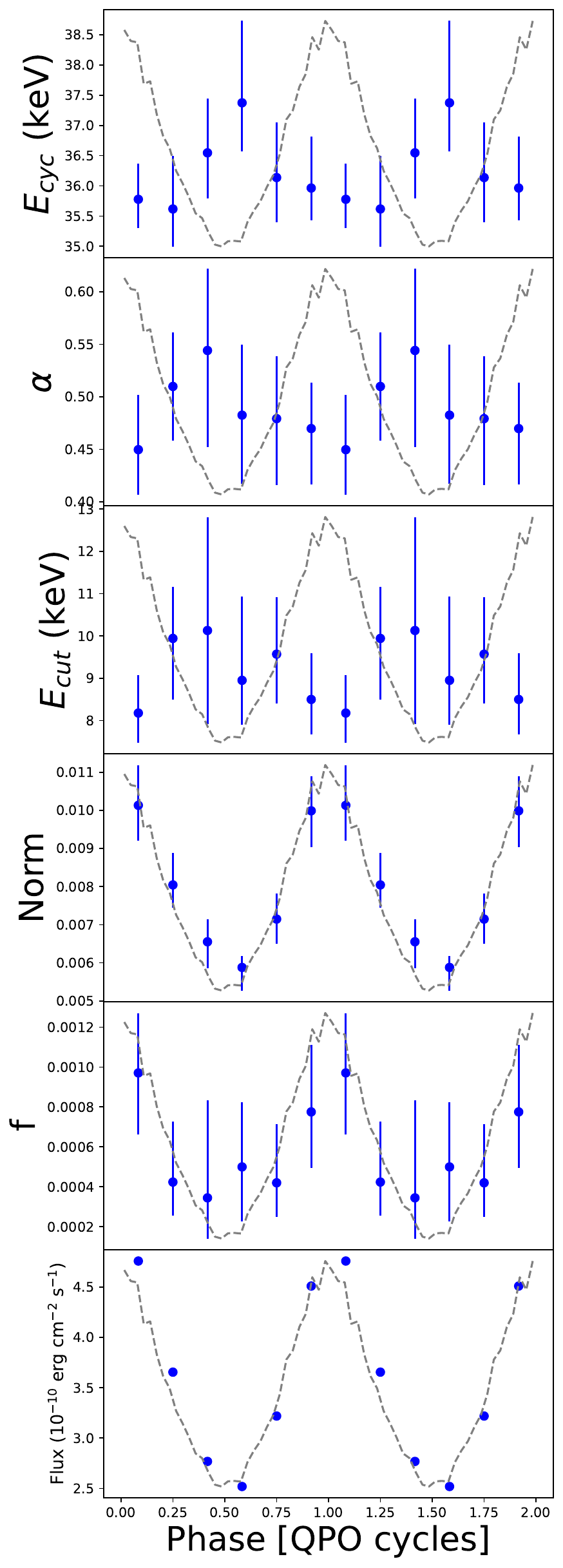}
    \hspace{0.05cm}
    \includegraphics[width=0.215\textwidth]{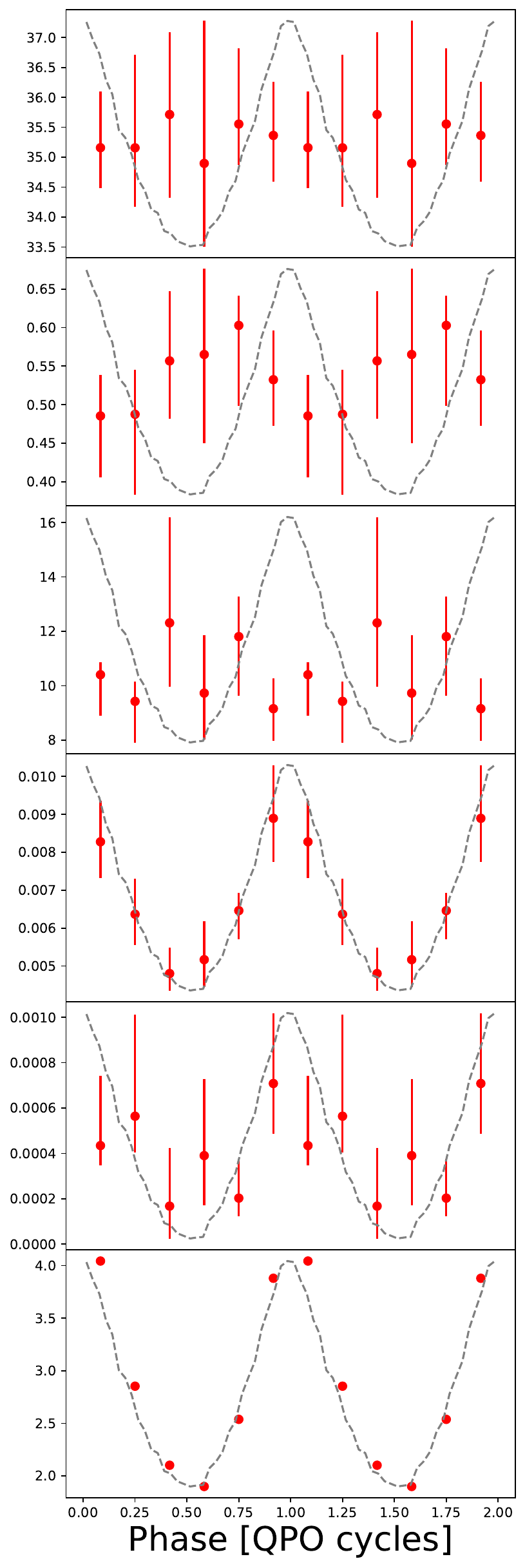}
    \hspace{0.05cm}
    \includegraphics[width=0.215\textwidth]{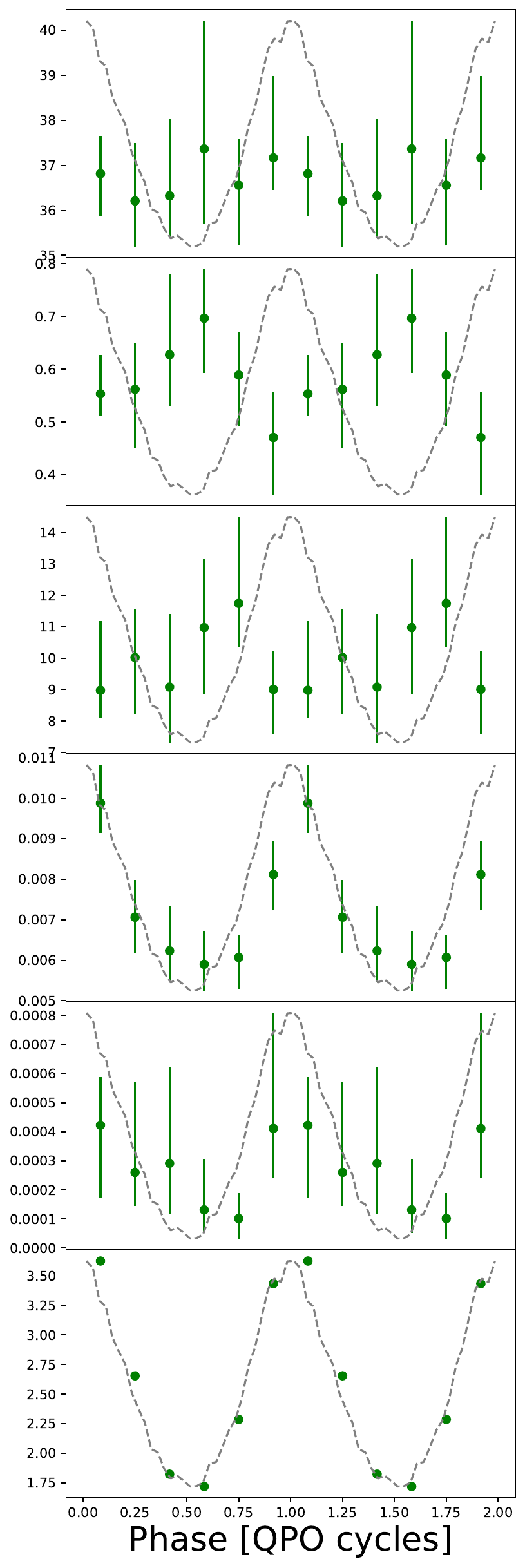}
    \hspace{0.05cm}
    \includegraphics[width=0.215\textwidth]{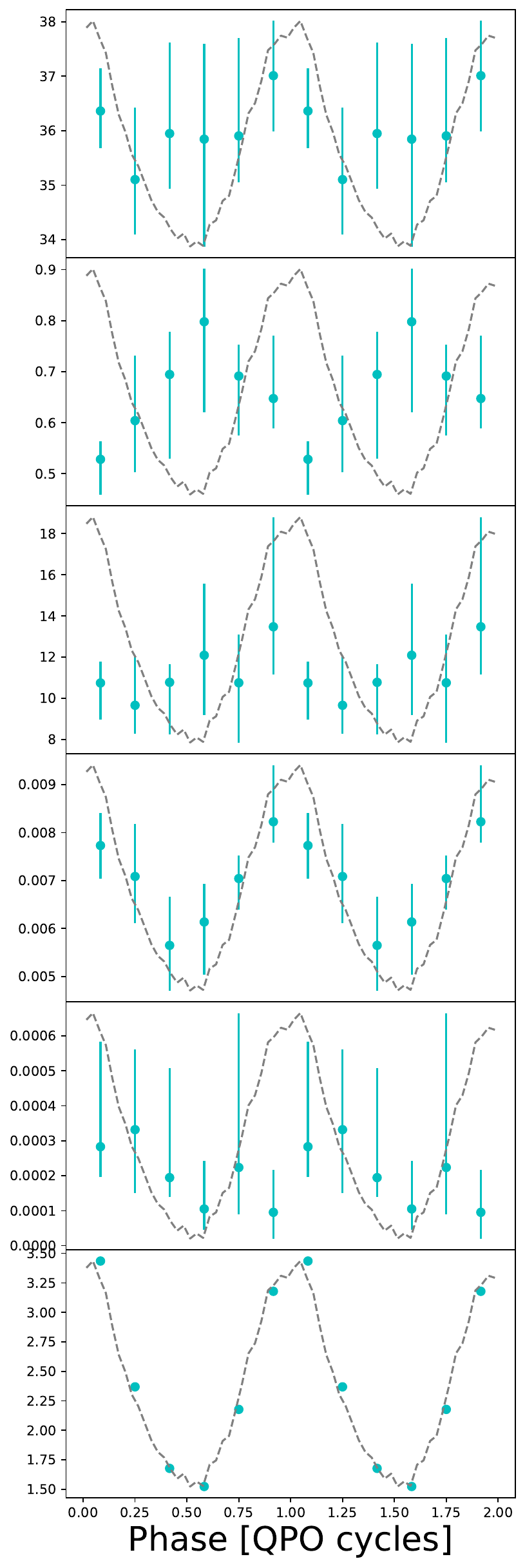}
    \caption{Results of the QPO-phase-resolved spectroscopy for four observations, i.e., 90901318002 (blue), 90901318006 (red), 90901318008 (green) and 90901318010(cyan). 
    Grey dashed lines represent QPO waveforms in the energy band of 4-79\, keV. 
    }
    \label{fig:phase-solved}
\end{figure*}

\begin{figure}[h!]
    \centering
    \includegraphics[width=0.45\textwidth]{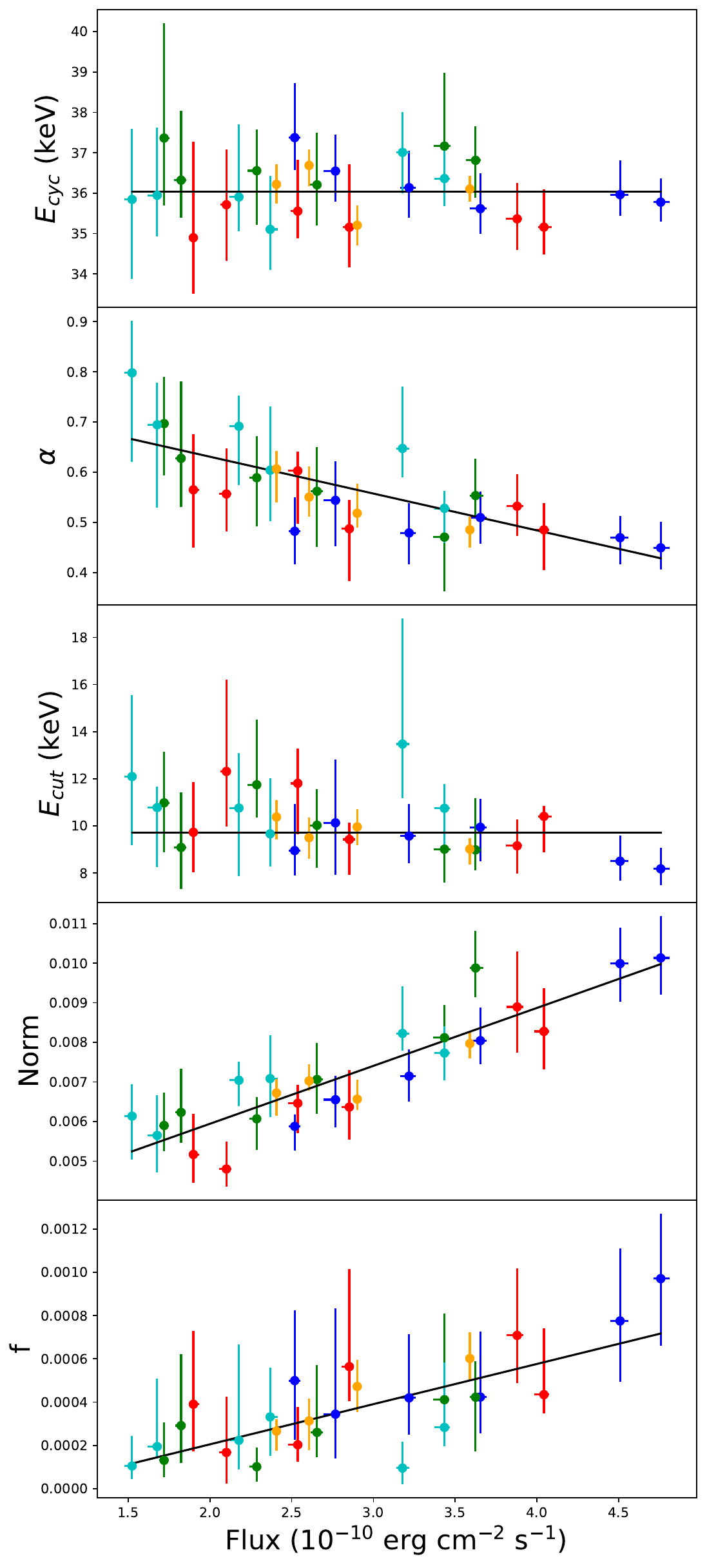}
    \caption{The evolution of spectral parameters with the observed flux in the 4--79 keV band. Orange points represent spectral parameters from the QPO-phase-averaged spectra.  Other points show QPO-phase-resolved results for four observations: 90901318002 (blue), 90901318006 (red), 90901318008 (green), and 90901318010 (cyan).
    For each observation, QPOs are divided into six phases as mentioned in Section~\ref{subsec:3.2}.
    Black lines are linear fits to the QPO-phase-resolved spectral parameters. }
    \label{fig:fit}
\end{figure}

According to QPO phases determined by the HHT, we could perform phase-resolved spectral analysis and investigate spectral evolution with QPO phases. 
In practice, we divided each observation into six QPO phases and extracted their corresponding spectra.
Then we fitted these phase-resolved spectra using the same model mentioned above.    
Due to poorer statistics, the width and depth of the fundamental CRSF and its harmonic cannot be well constrained.
Thus we set their values to the best-fit parameters obtained from phase-averaged spectra.
We show resulting spectral variations in Figure~\ref{fig:phase-solved}.
In all observations, the CRSF energy $E_{\rm cyc}$ and $E_{\rm cut}$ do not present significant changes for different QPO phases. 
On the other hand, the flux and normalizations $N$, $f$ exhibit similar modulations compared to the QPO waveform, where the flux was estimated in the energy range of 4-79\,keV (see Eq.~\ref{eq:npex} for definitions of $N$, $f$).
Conversely, the spectral index $\alpha$ is clearly anti-correlated with the flux, i.e., the spectrum is harder for a larger flux.


To describe the flux-dependent evolution, we performed linear fits for all parameters (Figure~\ref{fig:fit}).
We found that $E_{cyc}$ remains approximately constant at 36.0\,keV in the flux range of $(1.5-5.0)\times10^{-10}$\,$\rm erg\ cm^{-2}\ s^{-1}$.
In addition, we found that the spectral evolution with the QPO-modulated flux is well consistent with the parameters obtained from average spectra (yellow points) that vary at a much longer timescale.
This may suggest that spectral variations during QPOs and those detected among different observations have the same underlying reason, i.e., changes of the accretion rate.


\section{DISCUSSION} \label{sec:4}
QPOs have been widely detected in low-mass X-ray binaries hosting either neutron stars or black holes.
In this paper, we report their properties in a unique highly magnetized LMXB 4U 1626$-$67 during the recent spin-down episode.
Among four {\it NuSTAR} observations, we found that power spectra have a similar shape and a prominent QPO component presents around 47\,mHz, consistent with previous reports \citep[e.g.,][]{Sharma..2023MNRAS.526L..35S, Tobrej..2024MNRAS.528.3550T}.
Using the HHT method, we successfully constructed QPOs' waveforms, which appear as a sine-like modulation.
The QPO amplitude generally increases with energy, although this trend seems to be saturated at high energies (i.e., $E \gtrsim 15$\,keV; see Figure~\ref{fig:energy-resolved}).

This energy-dependence of QPOs' rms suggests spectral variations across different QPO phases.
Indeed, we found positive or negative correlations between spectral parameters and flux based on the QPO-phase-resolved spectroscopy. 
What is interesting is that these correlations are well in agreement with those obtained from QPO-averaged spectra (i.e., comparing yellow points with others in Figure~\ref{fig:fit}).
This suggests that the QPO modulation may reflect actual changes of the accretion rate, rather than the obscuration effect.
If this is the case, QPOs will offer an independent probe to study the short-term variability of the accretion rate.
We note that the HHT method presented in this paper is analogous to the traditional ``pulse-to-pulse" analysis, in which pulses of different intensities are divided into multiple groups \cite[e.g.,][]{Klochkov2011,Shui2024}.
Its advantage lies in its capability to decompose components in the frequency domain, even for short time scales when the statistical noise dominates and the instantaneous flux can not be accurately estimated in the time domain.
If QPOs are due to accretion rate changes, the evolving photon index with the QPO-modulated flux ($\alpha$ in Fig.~\ref{fig:fit}) is in agreement with the spectral hardness evolution found at low luminosities of six transient X-ray pulsars \citep{Postnov2015}.
This suggests that 4U 1626$-$67 remains in the sub-critical regime even around QPO peaks.
In addition, the centroid energy of cyclotron lines keeps constant over different QPO phase-resolved fluxes (Fig.~\ref{fig:fit}), implying that the line-forming region is always at a relatively low altitude, as expected in the sub-critical regime \citep[see Fig.~9 in][]{Staubert..2019A&A...622A..61S}.

On the other hand, the accretion rate is not the only factor affecting the properties of the observed emission.
For example, in many sources temporal and spectral properties are not exactly the same when comparing the rising and decay phases of outbursts for a given luminosity \citep[e.g.,][]{Doroshenko2017,Hu2023}.
The origin of this hysteresis effect is still unknown.
A plausible reason is changes of the accretion disk structure and its interaction with the magnetosphere of the neutron star \citep[see discussions in][]{Kong2021}.
In our study, we find that within each observation, the spin pulsed fractions are nearly independent of QPO phase, despite variations in flux. On the other hand, for a given QPO flux level, the pulsed fractions differ noticeably across different observations (Fig.~\ref{fig:flux_pulsed_fraction}, right panel).
This may suggest that the pulsed fraction is more sensitive to the disk structure than to the instantaneous accretion rate. When comparing different observations, the disk structure, e.g., the thickness of the flow outside the magnetosphere, might be different due to the pile-up of backflows (see below), leading to different pulsed fractions.

Although the formation mechanism of QPOs is poorly known, it is believed that they are probably attributed to interactions between the accretion matter and the NS magnetic field. Unlike the cases in wind accreting systems, such as Vela X-1, where QPOs may originate from the quasi-periodic structure in the stellar wind of the companion star \citep{Kreykenbohm2008}, QPOs in 4U 1626$-$67 are more likely caused by the variability of the accretion disk. In disk accretion systems, the QPO frequency are usually interpreted by either the Keplerian frequency ($\nu_{\rm k}$) at the inner disk radius \citep[][]{van..1987ApJ...316..411V} or the beat frequency between the neutron star spin frequency ($\nu_{\rm NS}$) and $\nu_{\rm k}$, i.e., $\nu_{\rm QPO} = |\nu_{\rm k} - \nu_{\rm NS}|$ \citep[][]{Alpar..1985Natur.316..239A}.
4U 1626$-$67 is unique among NS LMXBs due to its strong magnetic field, which truncates the accretion disk at the magnetospheric radius
\begin{equation}
R_{\rm m}=k\left(\frac{\mu^4}{2GM\dot{M}^2}\right)^{\frac{1}{7}},
\label{eq_Rm}
\end{equation}
where $\mu=\frac{1}{2}BR^3$ is the magnetic dipole moment, $M$ is the mass of the neutron star, $\dot{M}$ is the accretion rate, and $k=0.5$ for a disk accretion \citep{Ghosh1979}.
Assuming an isotropic radiation, the accretion rate can be estimated as $\dot{M}=\frac{4\pi D^2 F}{\eta c^2}$, where $\eta\approx0.2$, $D\sim (5-13)$\,kpc is the range of the distance to 4U 1626$-$67 \citep{Chakrabarty1998}, $F\sim(1.5-4.5)\times10^{-10}$\,$\rm erg\,cm^{-2}\,s^{-1}$ is the bolometric flux range calculated from the broadband spectral analysis.
Given a QPO frequency ($\nu_{\rm QPO}$) that equals the Keplerian frequency at $R_{\rm m}$, the inner disk radius should be $R_{\rm m}=(\frac{\nu_{\rm NS}}{\nu_{QPO}})^{2/3}R_{\rm co} \approx 2\,R_{\rm co}$, where $R_{\rm co}=(\frac{GM}{4\pi^2\nu_{\rm NS}^2})^{1/3}$ is the co-rotation radius where the local Keplerian frequency equals the neutron star spin frequency.
However, it is believed that when $R_{\rm m} \gg R_{\rm co}$, the magnetic dipole will act like a rotating ``propeller", preventing from the accretion process onto the neutron star \citep{Illarionov..1975A&A....39..185I}. 
Under this assumption, most of the mass in the disk is expelled as an outflow, which conflicts with the observed accretion with QPOs. Thus the Keplerian frequency model can be ruled out.

\begin{figure}
    \centering
    \includegraphics[width=0.45\textwidth]{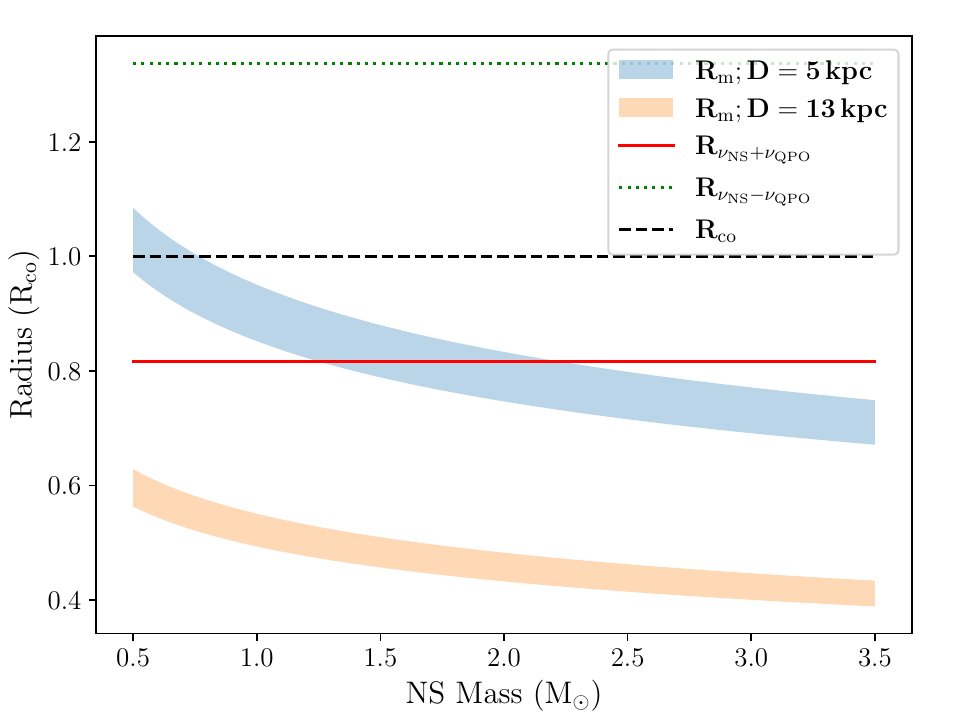}
    \caption{
    The magnetospheric radius $R_{\rm m}$ calculated from Eq.~\ref{eq_Rm} assuming $k=0.5$.
    Different shadows indicate the distance range (5-13\,kpc) we considered in this paper.
    The shadow widths correspond to the flux range (1.5-4.5 $\rm \times 10^{-10}\,erg\,cm^{-2}\,s^{-1}$) calculated from the broadband spectral analysis.
    The solid and dot lines present two radii with the local Keplerian frequency $\nu_{\rm k} = \nu_{\rm NS} + \nu_{\rm QPO}$ and $\nu_{\rm k} = \nu_{\rm NS} - \nu_{\rm QPO}$ based on the beat frequency model, and their intersections with shadows are possible solutions.
    }
    \label{fig:M_R}
\end{figure}

\begin{figure*}
    \centering
    \includegraphics[width=0.7\textwidth]{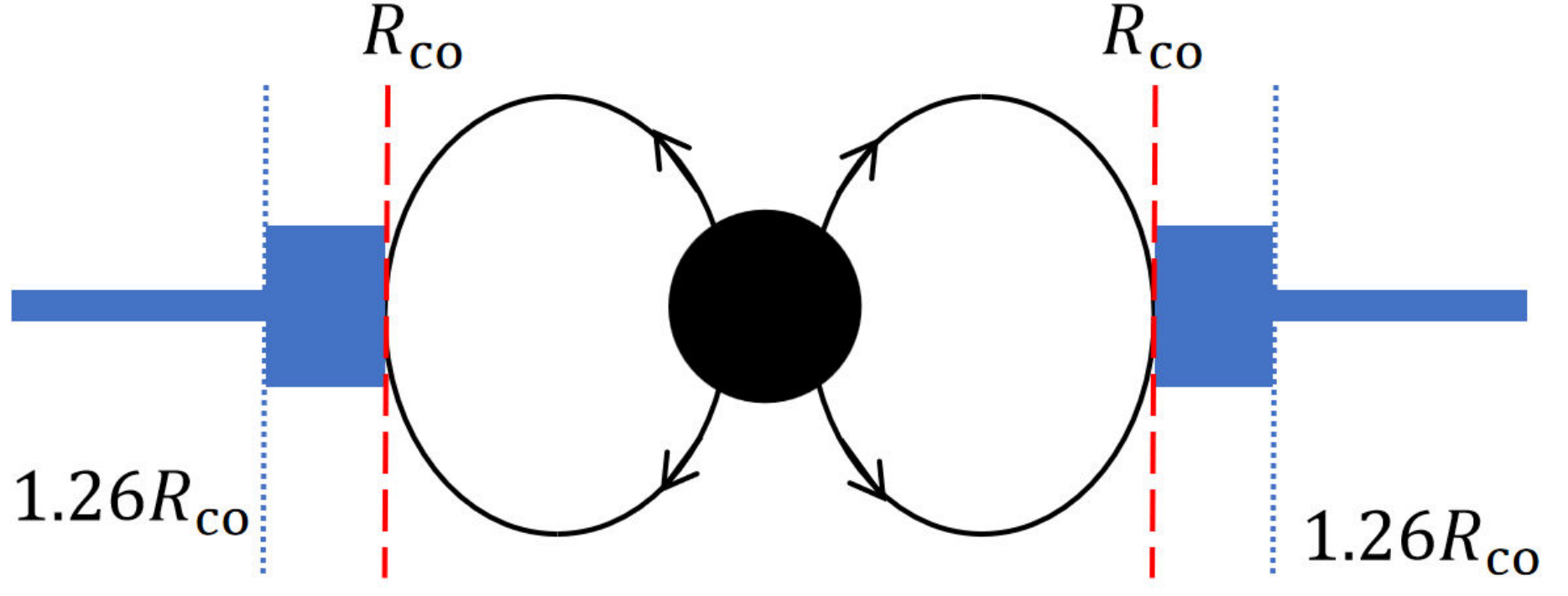}
    \caption{
    A possible configuration of 4U 1626$-$67, where black lines represent the magnetosphere. The red dashed line marks the co-rotational radius and the blue doted line is the radius above which the strong propeller effect works.
    Outside the magnetospheric radius, there might be a region with a trapped disk influenced by the pile-up of backflows. 
    }
    \label{fig:sketch}
\end{figure*}

On the other hand, when considering beat frequency models, the $R_{\rm m}$ is expected to be $(\frac{\nu_{\rm NS}}{\nu_{\rm NS} + \nu_{\rm QPO}})^{2/3}R_{\rm co} \approx 0.8\,R_{\rm co}$ or $(\frac{\nu_{\rm NS}}{\nu_{\rm NS} - \nu_{\rm QPO}})^{2/3}R_{\rm co}\approx 1.3\,R_{\rm co}$.
If we consider the ``plus" solution, i.e., $R_{\rm m} \approx 0.8\,R_{\rm co}$, a possible parameter space is shown in Fig.~\ref{fig:M_R}.
The mass of the neutron star is expected to be $\sim(1.5-2)$$M_{\odot}$ when assuming a distance of 5\,kpc, while a larger distance (e.g., 13\,kpc) can be ruled out.
In this ``plus" scenario, the magnetosphere interacts with a thin accretion disk, whose inner radius will move inwards at a higher accretion rate, leading to a higher QPO frequency.
In the literature, this positive correlation between the QPO frequency and the luminosity have been reported in a few sources \citep{Angelini1989, Finger1996, Ma2022}. 
However, the QPO frequency of 4U 1626$-$67 appears consistently around 47\,mHz in our study and previous reports \citep{Tobrej..2024MNRAS.528.3550T}, independent of the luminosity.
This challenges theoretical models which are based on the inner disk radius, such as the aforementioned beat frequency model and  the precession disk model proposed by \citet{Shirakawa..2002ApJ...565.1134S}.
In addition, this solution can not explain why QPOs in this source are prominent only during the spin-down state \citep{Jain..2010MNRAS.403..920J}.

In the following, we discuss the ``minus" solution in the beat-frequency model mentioned above.
Conventionally, only the ``plus" solution is allowed to avoid the ``propeller" regime, while recent studies suggest that the ``minus'' solution is still possible with accretion in some cases \citep[for details, see e.g.,][]{DAngelo2012,Ertan2021, Gencali2022}.
This is due to the fact that a significant propeller happens only if the speed of the rotating magnetic line exceeds the local escape speed, corresponding to the region outside 1.26\,$R_{\rm co}$.
While for a magnetospheric radius between $R_{\rm co}$ and 1.26\,$R_{\rm co}$, the launched outflow will return back to the disk, referred to as the weak propeller regime.
As a result, a trapped disk may form as a boundary region due to the pile-up of the backflow with an inner radius close to $R_{\rm co}$.
A simplified sketch of the discussed configuration is presented in Fig.~\ref{fig:sketch}.
We speculate that QPOs in 4U 1626$-$67 might be generated by the beat frequency model, because the location (i.e., around $1.3\,R_{\rm co}$) of the local Kepler frequency required in the ``minus" solution coincidentally corresponds to the outer edge of the boundary region.
In addition, the boundary region is almost luminosity-independent.
This can naturally explain why the QPO frequency keeps stable at different luminosities. In the literature, luminosity-independent QPO frequencies were also reported in Cen~X-3, 4U~0115+63 and Her~X-1 \citep{Liu..2022MNRAS.516.5579L, Rm_Rco..2021MNRAS.503.6045D,HerX1QPO..2025ApJ...980..194Y} and explained by an instability when the accretion disk is truncated near the corotation radius, similar to our scenario.
In addition, the ``minus" solution implies a weak propeller state, in which the spin-down torque would be significant. 
This is in good agreement with the fact that QPOs are associated with the spin evolution of the source.

\begin{acknowledgments}
This work is supported by the National Natural Science Foundation of China under grants No. 12173103 and 12261141691. 
\end{acknowledgments}

\appendix
\renewcommand\thefigure{\thesection.\arabic{figure}}    
\setcounter{figure}{0}  
\section{Variational Mode Decomposition}\label{sec:appendix A}
VMD is an adaptive, non-recursive technique that decomposes a signal into a set of intrinsic mode functions \citep{Dragomiretskiy..2014ITSP...62..531D}. It aims to identify several narrow-bandwidth signal components whose sum can best reconstruct the original signal. This method offers improved noise robustness and mode separation compared to the original empirical mode decomposition \citep{Huang..2008RvGeo..46.2006H}. The core of the VMD method is solving an optimization problem to decompose a light curve into several IMFs: 
\begin{equation}
\min_{\{u_{\rm k}\}, \{\omega_{\rm k}\}} \left\{ a \sum_{{\rm k}=1}^{K} BW_{\rm k}^2 + \left\| f(t) - \sum_{{\rm k}=1}^{K} u_{\rm k}(t) \right\|_2^2 \right\},
\label{eq:vmd}
\end{equation}
where \textit{a} is the bandwidth limitation parameter and $K$ is the number of IMFs. 
$f(t)$ and $u_{\rm k}(t)$ represent the raw light curve and the contribution from the k-th IMF.
$\omega_{\rm k}$ represents the center frequency of the k-th IMF and $BW_{\rm k}$ is its bandwidth.
$f(t)$, $u_{\rm k}(t)$, $\omega_{\rm k}$, and $BW_{\rm k}$ can be solved iteratively by the algorithm, while \textit{a} and $K$ are two input parameters.

The IMFs are nearly orthogonal in the frequency domain and each IMF is narrow-band around its center frequency. In practice, Vmdpy allows for the configuration of several distinct parameters to perform VMD. In this work, we set \textit{DC}=0, \textit{tau}=0, and \textit{init}=2, along with the values of \textit{K} and \textit{a} as listed in Table \ref{tab:observations}. Here, \textit{DC}=0 indicates that the first mode is not constrained to be a direct current (zero-frequency) component, allowing it to capture the most dominant low-frequency fluctuation freely. \textit{tau}=0 enables the noise-tolerance mode, emphasizing robust decomposition in the presence of noise rather than strict fidelity to the original signal. \textit{init}=2 initializes the center frequencies of the modes randomly, facilitating an adaptive search for frequency bands without presupposing their distribution.

\begin{table}
    \centering
    \caption{Best-fit spectral parameters of 4U 1626$-$67 in four observations using the \textit{const$*$tbabs$*$cyclabs$*$(npex+gaussian)} model. 
    We assumed a distance of 10\,kpc to calculate the luminosity.
    }
    \label{tab:fitting para}
    \begin{tabular}{lllllll}
        \hline
        Model & Parameters & Units & 90901318002 & 90901318008 & 90901318006 & 90901318010 \\
        \hline
        NPEX & $E_{\rm cut}$ & keV & 9.01$^{+0.65}_{-0.45}$ & 9.50$^{+0.86}_{-0.89}$ & 9.96$^{+0.79}_{-0.75}$ & 10.37$^{+0.94}_{-0.71}$ \\
        & $\alpha$ &  & 0.49$^{+0.03}_{-0.03}$ & 0.55$^{+0.04}_{-0.06}$ & 0.52$^{+0.03}_{-0.06}$ & 0.61$^{+0.07}_{-0.04}$ \\
        & f & 10$^{-4}$ & 6.02$^{+1.05}_{-1.22}$ & 3.13$^{+1.36}_{-1.02}$ & 4.72$^{+1.17}_{-1.24}$ & 2.66$^{+0.91}_{-0.56}$ \\
        & Norm & 10$^{-3}$ & 7.97$^{+0.37}_{-0.31}$ & 7.02$^{+0.25}_{-0.42}$ & 6.56$^{+0.27}_{-0.49}$ & 6.72$^{+0.57}_{-0.36}$ \\
        $C_{\text{FPMB/FPMA}}$ & Factor & & 1.00$^{+0.01}_{-0.01}$ & 1.01$^{+0.01}_{-0.01}$ & 1.00$^{+0.01}_{-0.01}$ & 0.99$^{+0.01}_{-0.01}$ \\
        Cyclabs & $E_{cyc}$ & keV & 36.10$^{+0.31}_{-0.33}$ & 36.69$^{+0.52}_{-0.39}$ & 35.21$^{+0.51}_{-0.48}$ & 36.22$^{+0.48}_{-0.50}$ \\
        & Depth &  & 2.42$^{+0.17}_{-0.14}$ & 2.13$^{+0.20}_{-0.13}$ & 2.26$^{+0.18}_{-0.23}$ & 2.60$^{+0.28}_{-0.26}$ \\
        & Width & keV & 6.45$^{+0.86}_{-0.68}$ & 6.10$^{+0.89}_{-0.86}$ & 7.96$^{+0.89}_{-0.93}$ & 5.80$^{+0.83}_{-0.46}$ \\
        \hline
        Luminosity& &$10^{36} \, \text{erg}\,\text{s}^{-1}$ & 4.29$^{+0.02}_{-0.02}$ & 3.12$^{+0.01}_{-0.03}$ & 3.47$^{+0.01}_{-0.03}$ & 2.88$^{+0.03}_{-0.02}$  \\
        $\chi^2$ / d.o.f. & & & 265/251 & 308/233 & 248/228 & 199/218 \\
        \hline
    \end{tabular}
\end{table}

\section{Comparing different combinations of $K$ and $\lowercase{a}$} \label{sec:appendix B}
\begin{figure*}[h!]
    \centering
    \includegraphics[width=0.45\textwidth]{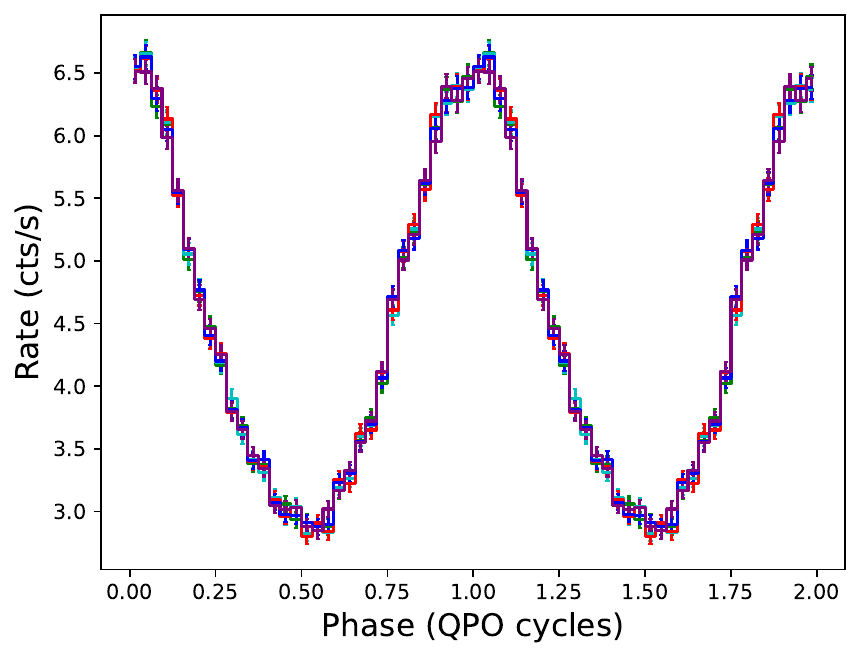}
    \hspace{0.1cm}
    \includegraphics[width=0.45\textwidth]{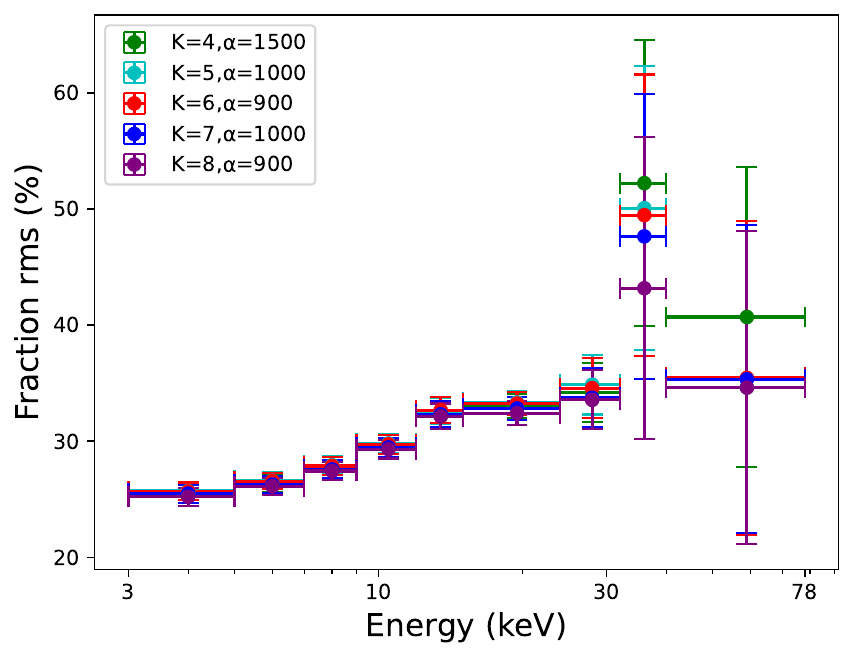}
    \caption{Left: reconstructed QPO waveforms by the VMD plus HHT technique assuming different $K$ and \textit{a} parameters.
    The data is adopted from the observational ID 90901318010 in the energy range of 4-79\,keV. 
    Right: the evolution of the fractional rms with energy, considering different $K$ and \textit{a} combinations.
    }
    \label{fig:vmd}
\end{figure*}

As outlined in Sect.~\ref{subsec:3.1}, the decomposition is somehow artificial, because the QPO in a power spectrum can be technically reconstructed using different combinations of $K$ and \textit{a}.
This is due to the fact that both parameters respond to the QPO width, which leads to an inevitable degeneracy.
We tested different combinations of $K$ and \textit{a} to decompose the raw lightcurve. The resulting modulation waveforms are quite similar, and the evolution of the fractional rms with energy presents a similar trend as well (Figure~\ref{fig:vmd}).
This suggests that a slightly different choice of $K$ and \textit{a} has only little influence on the energy-dependent waveform and therefore the QPO-resolved spectroscopy.

\section{Comparison of rms calculations}\label{sec:appendix C}
\begin{figure*}[h!]
    \centering
    \includegraphics[width=0.48\linewidth]{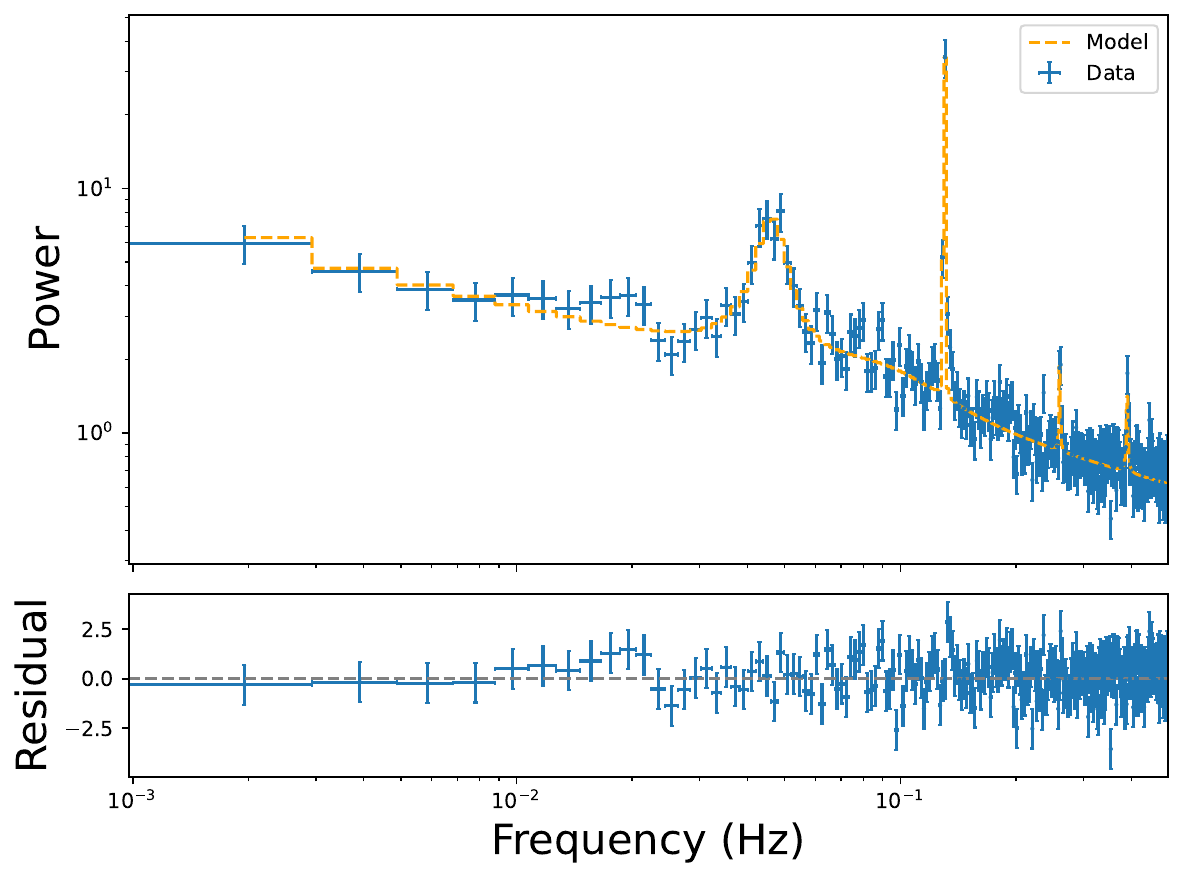}
    \includegraphics[width=0.48\linewidth]{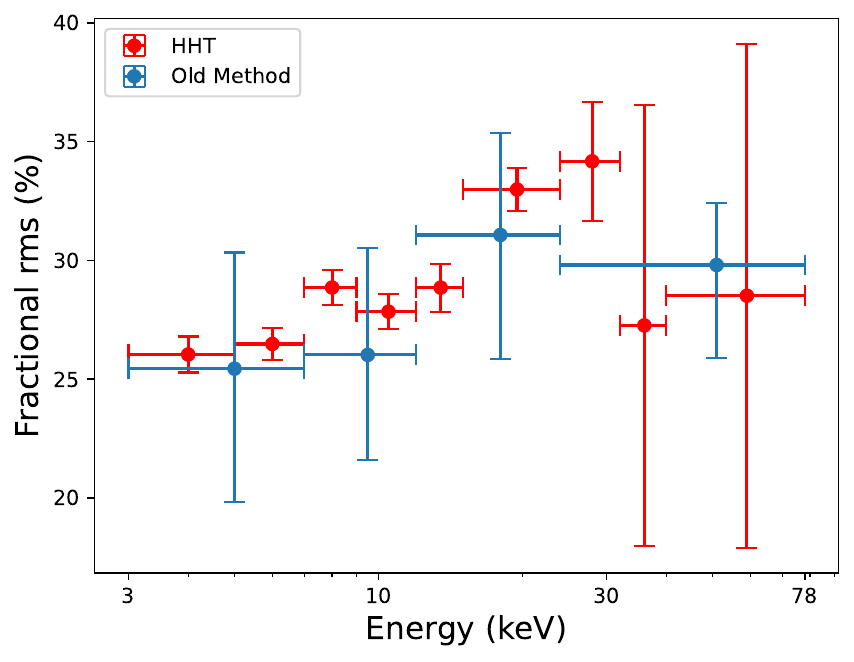}
    \caption{Left: a representative PDS of observation 90901318006 where the yellow dot line is the model for the fitting. Right: fractional rms values calculated by using the HHT (red points) and the conventional method based on PDSs (blue points).}
    \label{fig:LOR}
\end{figure*}

The rms of QPOs is conventionally estimated from power spectra \citep[for a review, see, e.g.,][]{Uttley2014}.
In this section, we compare fractional rms values derived from the HHT with those obtained from the old method. 
For the latter, we made calculations using the software {\tt Stingray}.
In practice, we modeled PDSs with a combination of one \textit{powerlaw} component representing the broadband noise and five \textit{lorentzian} components accounting for the QPO fundamental, its harmonic, and three coherent pulsations, respectively.
If a PDS was calculated using the Miyamoto normalization \citep{Belloni1990,Miyamoto1992}, the fractional rms of a QPO could be obtained by integrating this component over a frequency range and taking the square root.
In Figure~\ref{fig:LOR}, we show the results of ObsID 90901318006 as a representative example to make the comparison.
We divided the full energy band into four sub-bands, i.e., 3–7\,keV, 7–12\,keV, 12–24\,keV and 24–78\,keV, in order to investigate the energy-dependent fractional rms.
In the end, we find that the rms values estimated from the conventional method and the HHT as proposed in this paper are well consistent with each other. 
In general, the HHT method will provide smaller errors. 
But we caution that this is due to the fact that when calculating the rms of one energy band, the HHT considers the information from all energies which will somehow lead to the non-independence between different energy bands.


\bibliography{sample631}{}
\bibliographystyle{aasjournal}

\end{document}